\documentclass[twocolumn,aps]{revtex4-1}

\pdfpageattr {/Group << /S /Transparency /I true /CS /DeviceRGB>>}

\usepackage{graphicx} 
\usepackage{overpic}

\DeclareSymbolFont{lettersA}{U}{txmia}{m}{it}
\DeclareMathSymbol{\real}{\mathord}{lettersA}{"92}
\DeclareMathSymbol{\field}{\mathord}{lettersA}{"83}

\begin{document}

\title{Coping with qubit leakage in topological codes}

\author{Austin G. Fowler$^{1,2}$}

\affiliation{$^1$Department of Physics, University of California, Santa Barbara, California 93106,
USA \\
$^2$Centre for Quantum Computation and Communication
Technology, School of Physics, The University of Melbourne, Victoria
3010, Australia}

\date{\today}

\begin{abstract}
Many physical systems considered promising qubit candidates are not, in fact, two-level systems. Such systems can leak out of the preferred computational states, leading to errors on any qubits that interact with leaked qubits. Without specific methods of dealing with leakage, long-lived leakage can lead to time-correlated errors. We study the impact of such time-correlated errors on topological quantum error correction codes, which are considered highly practical codes, using the repetition code as a representative case study. We show that, under physically reasonable assumptions, a threshold error rate still exists, however performance is significantly degraded. We then describe simple additional quantum circuitry that, when included in the error detection cycle, restores performance to acceptable levels.
\end{abstract}

\maketitle

The threshold theorem of quantum computation states that a finite quantum gate error rate exists below which arbitrarily reliable quantum computation can be performed with only polylogarithmic overhead. The theoretical viability of quantum computation rests on this theorem \cite{Knil96b,Ahar97,Kita97}. Many different assumptions can be made, and different versions of the theorem proved, however typically it is assumed that qubits do not leak to non-computational states. Leakage has only been formally considered \cite{Alif07d} in a threshold theorem for concatenated quantum error correction (QEC) codes \cite{Shor95,Cald95,Stea96,Knil04c,Baco06}.

It is increasingly thought that topological QEC (TQEC) \cite{Brav98,Denn02,Bomb06,Raus07,Raus07d,Fowl09,Ohze09b,Katz10,Bomb10,Fowl11,Fowl12f} is more experimentally feasible than concatenated QEC. However, even recent TQEC threshold theorem proofs \cite{Harr04,Brav11,Fowl12e} assume leakage does not occur. It is therefore important to demonstrate that leakage, which is suffered by trapped ions \cite{Cira95}, quantum dots \cite{Byrd05,Fong11}, superconducting qubits \cite{Zhou05,Tao06,Motz09,Ferr10,Herr12b,Ghos13}, anyons \cite{Xu02,Ains11}, and many other systems, does not void the threshold theorem of TQEC, and that there are low-overhead and effective methods of handling leakage within TQEC.

The discussion is organized as follows. In Section~\ref{rep}, we show how the repetition code can be viewed as a special case of the surface code \cite{Fowl12f}, and present baseline repetition code simulations without leakage. In Section~\ref{leak}, we describe our stochastic model of leakage, and present and discuss repetition code simulations with leakage. In Section~\ref{circ}, we describe a simple circuit removing leakage, and present and discuss simulations with this circuitry included. Section~\ref{conc} concludes.

\section{The repetition code}
\label{rep}

\begin{figure}
\begin{center}
\resizebox{60mm}{!}{\includegraphics{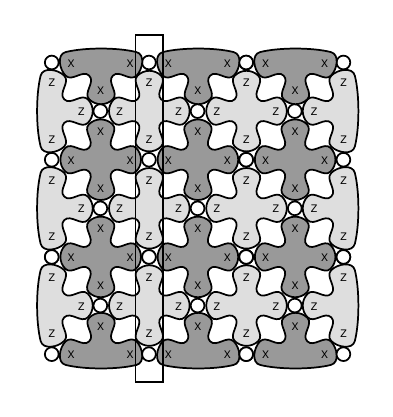}}
\end{center}
\caption{Distance 4 surface code. Rectangle indicates a vertical slice of the surface code that has the structure of a repetition code. Repetition code failure is thus a good and simple model of failure within the surface code.}\label{sc}
\end{figure}

Fig.~\ref{sc} shows a standard surface code. The 2-D structure of the surface code enables it to suppress both logical $X$ and logical $Z$ errors, however for most purposes it is sufficient to study the suppression of only logical $X$ or logical $Z$ errors due to the symmetry of the code. When studying the generic properties of a single type of logical error, it is sufficient to consider just a single vertical strip of qubits as additional width just introduces additional logical error pathways, but does not fundamentally change the need for a significant number of errors effecting at least half of a line of qubits from top boundary to bottom boundary. In this work, we shall therefore focus on the repetition code, which is precisely a code with $ZZ$ stabilizer generators on neighboring pairs of qubits in a line. This will enable us to focus the discussion on the unique properties of leakage, without unnecessarily getting bogged down in the details of the surface code.

\begin{figure}
\begin{center}
\resizebox{60mm}{!}{\includegraphics{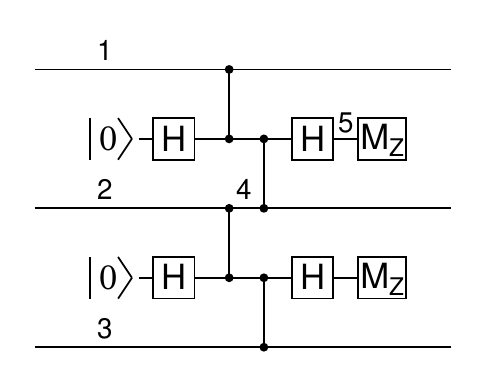}}
\end{center}
\caption{A single round of repetition code error detection. Circuit is for a distance three code, corresponding to three data qubits (the three horizontal lines). The code distance can be increased by repeating the vertical pattern, adding additional data qubits. Rounds of error detection repeat indefinitely.}\label{rep_circ}
\end{figure}

Fig.~\ref{rep_circ} shows the quantum circuit required to perform repetition code error detection. In the absence of errors, each measurement reports 0 if the neighboring data qubits are in the +1 eigenstate of $ZZ$ and 1 if they are in the -1 eigenstate. Referring to Fig.~\ref{rep_circ}, a location 1 error will lead to the top measurement differing from its previous value. This change is called a detection event \cite{Fowl12d}. A location 2 error will lead to two simultaneous detection events. A location 3 error will lead to a single detection event associated with the lower measurement qubit. Location 4 errors are more complex, with an immediate detection event associated with the top measurement qubit but the detection event associated with the lower measurement qubit not occurring until the next round of error detection. An erroneous measurement, location 5, will lead to two sequential measurement value changes, namely 010 or 101, meaning a pair of sequential detection events. All possible errors fall into one of these five geometric classes.

\begin{figure}
\begin{center}
\resizebox{80mm}{!}{\includegraphics{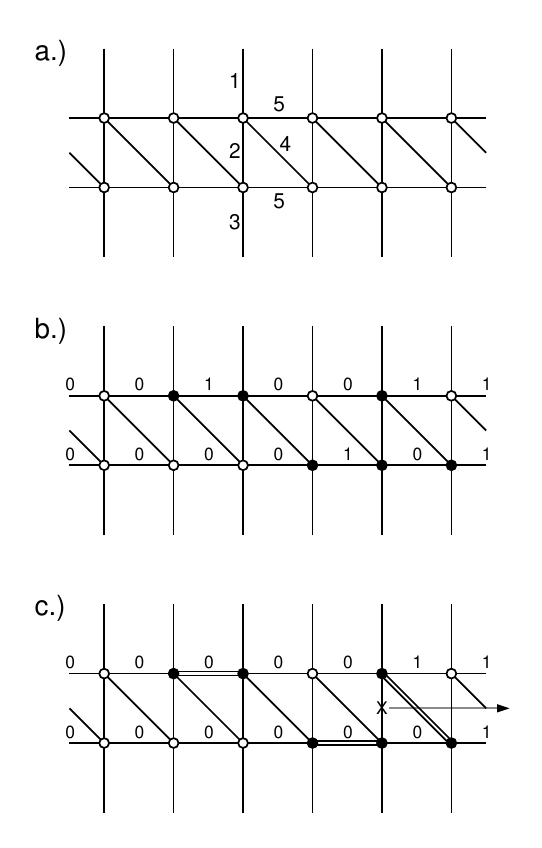}}
\end{center}
\caption{Time runs from left to right. a.) Geometric structure of classes (lines) of single errors with the potential to lead to detection events at their endpoints. Each layer of the periodic structure corresponds to a round of error detection. b.) Specific sequence of measurement results leading to specific detection events (filled circles). c.) Minimum weight perfect matching of the detection events and correction of two classical measurement results and insertion of a Pauli frame $X$ operator.}\label{problem}
\end{figure}

Each geometric class of potential error can be represented as a line connecting a pair of dots at the space-time locations of potential detection events as shown in Fig.~\ref{problem}a. Each line is given a weight proportional to $-$$\ln p$, where $p$ is the total probability of all errors leading to detection events at the endpoints of that line. This means that all line weights are strictly positive, and low probability lines are associated with high weights.

Fig.~\ref{problem}b shows a possible sequence of measurement values and the associated inferred pattern of detection events. Given dots, lines, and detection events, the minimum weight perfect matching algorithm \cite{Edmo65a,Edmo65b,Fowl11b,Fowl12c} can be used to find a set of paths of lines that connects all detection events in pairs or individually with a boundary such that the total weight of all lines in the set is minimal. This corresponds to a high probability pattern of errors leading to the observed detection events.

Fig.~\ref{problem}c shows a minimum weight perfect matching of Fig.~\ref{problem}b. The two class 5 matched lines result in a classical bit-flip each. The class 4 matched line results in the belief that an $X$ error must now be present on the central data qubit. If all data qubits were now measured and the measurement string 010 returned, our belief that an $X$ error is present on the central qubit would mean that we interpret this string as 000. If we had initially prepared our data qubits in 000, this would imply successful storage.

\begin{figure}
\begin{center}
\resizebox{85mm}{!}{\includegraphics[viewport=60 60 545 430, clip=true]{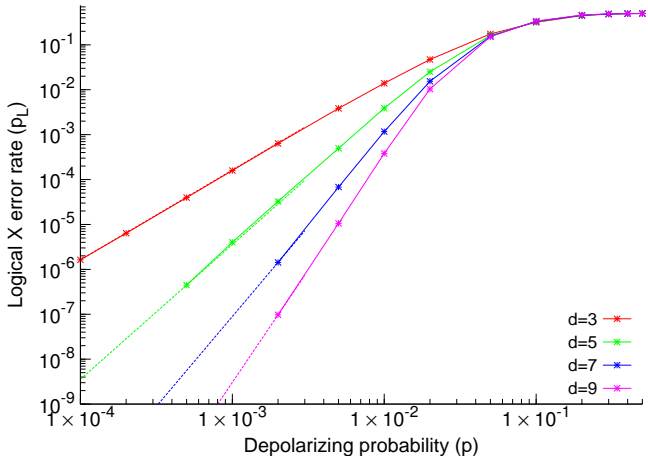}}
\end{center}
\caption{(Color online) probability of logical $X$ error $p_L$ as a function of the depolarizing error probability $p$ for distances $d=3,5,7,9$. Referring to the left of the figure, the distance $d=3,5,7,9$ curves are ordered top to bottom. Quadratic, cubic, quartic, and quintic curves have been fit through the lowest data points, respectively.}\label{logx}
\end{figure}

In simulations, we can perform many rounds of error detection and regularly check whether logical errors have occurred, thus calculating the probability of logical error per round of error detection. In this work we assume all gates, namely initialization, measurement, identity, Hadamard, and CZ, have equal depolarizing error rate $p$. Our open source simulation software Autotune \cite{Fowl12d} can also handle arbitrary asymmetric Pauli error models for each gate. Fig.~\ref{logx} shows the probability of logical $X$ error $p_L$ as a function of $p$ for a range of code distances $d$. The code distance $d$ of a repetition code is simply the number of data qubits used. In theory, a distance $d$ code should only fail if at least $\lceil d/2 \rceil$ errors occur. This can be observed with the low $p$ asymptotic forms of the $d=3,5,7,9$ curves being quadratic, cubic, quartic, and quintic, respectively.

\section{Simulating leakage}
\label{leak}

The simulations of the previous section included only the qubit error states $I$, $X$, $Y$, $Z$. We now wish to extend our simulations to include leakage $L$. We shall model leakage as only potentially occurring after Hadamard and CZ as in many quantum technologies initialization, measurement, and identity do not have the potential to cause leakage. It would be straightforward to associate leakage with other gates, if deemed appropriate. We shall set the probability of leakage per Hadamard and CZ to be $0.1p$, which we consider a high probability of leakage. The two qubits involved in a CZ gate shall each have independent probability $0.1p$ of leakage.

After a qubit leaks, further errors $X$, $Y$, $Z$, $L$ will be modeled as leaving the qubit in $L$. We model the decay of leaked states back to computational states with a fixed 1\% probability per identity, Hadamard, and CZ gate of decaying from $L$ to a randomly chosen $I$, $X$, $Y$, $Z$ error state. Measurement of a qubit in $L$ shall give a random classical result with no indication that leakage had occurred.

\begin{figure}
\begin{center}
\resizebox{85mm}{!}{\includegraphics[viewport=60 60 545 430, clip=true]{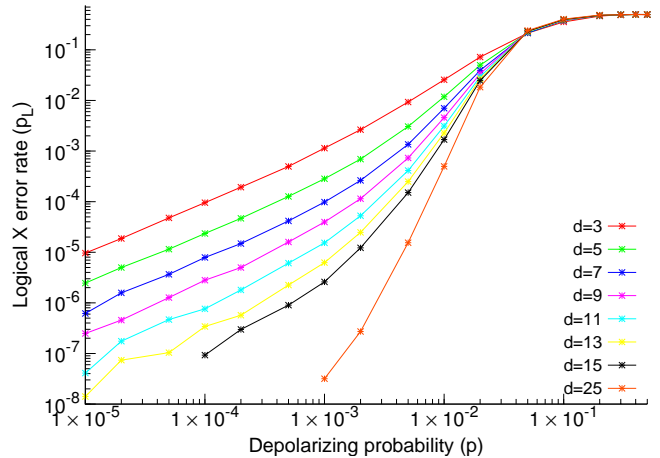}}
\end{center}
\caption{(Color online) probability of logical $X$ error $p_L$ as a function of the depolarizing error probability $p$ and leakage error probability $0.1p$ for a range of distances $d=3\ldots 25$. Referring to the left of the figure, the distance increases top to bottom.}\label{logx_loss}
\end{figure}

When CZ occurs between a qubit in $L$ and a qubit not in $L$, the latter will be scrambled to a randomly chosen $I$, $X$, $Y$, $Z$ error state. This severe model of leakage can be made more severe by decreasing the decay probability, making leakage longer lived. Note that it is important that leakage errors do not propagate to other qubits as this would lead to a catastrophic cascade of leakage errors ending only when all qubits in the computer had leaked. Fig.~\ref{logx_loss} shows the performance of the repetition code with finite lifetime leakage as described above.

\begin{figure}
\begin{center}
\resizebox{80mm}{!}{\includegraphics{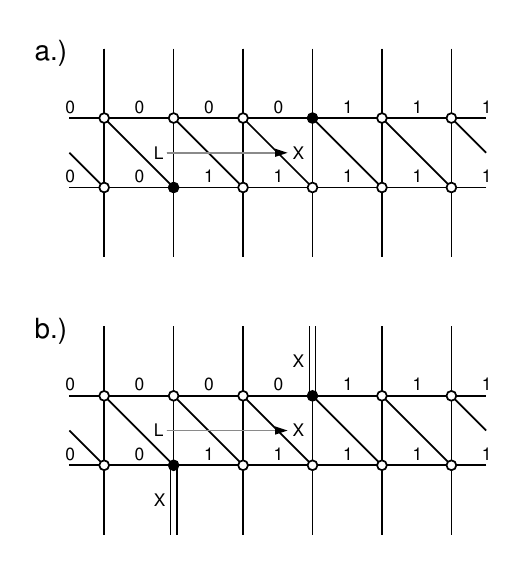}}
\end{center}
\caption{a.) Example of a leakage event that decays to an $X$ error. During the lifetime of the leakage event, neighboring measurement qubit results are random, with the indicated results perfectly possible. This pattern of measurement results leads to two detection events separated by a significant amount of time. b.) Minimum weight perfect matching infers that two individual $X$ errors are likely, leading to a logical error.}\label{example}
\end{figure}

It is immediately apparent from Fig.~\ref{logx_loss} that leakage has severely degraded the performance of the repetition code. The reason for this is simple --- a single leakage event has the potential to cause a logical error. Consider leakage of the central data qubit in Fig.~\ref{rep_circ}. Each time the neighboring measurement qubits interact with this data qubit, they will be randomized, leading to random measurement values. Suppose after a period of time the leaked data qubit decays to an error state $X$. Fig.~\ref{example} shows how randomized measurement values can lead to a logical error in a distance 3 code from a single leakage event. Note critically that this pattern of detection events could just have easily have been generated by two data qubit errors, and there is no way to distinguish between these two cases with the available information.

Similar random patterns of measurement values can lead to logical errors in higher distance codes, however longer and longer chains of specific measurement values are required and these chains become increasingly unlikely, resulting in logical error suppression with increasing code distance, preserving the existence of a threshold error rate. The specific pattern of random measurements shown in Fig.~\ref{example} could be made correctable by inserting some longer range lines, enabling the detection events to be matched with low weight, however this would lead to a logical error if these two detection events were in fact generated by two data qubit errors. Furthermore, no matter how the matching problem is restructured, there will always be random patterns of measurements arising from single leakage events that lead to logical failure, resulting in poor performance unless specific hardware steps are taken to regularly remove leakage from the system.

\section{Suppressing leakage}
\label{circ}

\begin{figure}
\begin{center}
\resizebox{50mm}{!}{\includegraphics{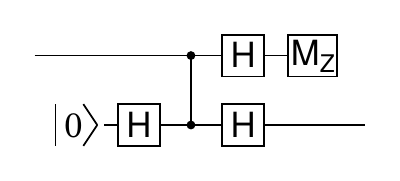}}
\end{center}
\caption{Circuit teleporting the top qubit to the bottom qubit, converting leakage errors on the top qubit to Pauli errors on the bottom qubit.}\label{leak_recover}
\end{figure}

Given leakage of the form described in the previous Section, the simple circuitry shown in Fig.~\ref{leak_recover} can be used to convert leakage errors $L$ into standard Pauli errors $I$, $X$, $Y$, $Z$ which can then be handled with regular error correction. In a distance 3 code, which typically has three data qubits $q_1q_3q_5$ and two measurement qubits $q_2q_4$, we can add an additional qubit $q_6$ and after the standard error detection circuit of Fig.~\ref{rep_circ} teleport each data qubit to the qubit below, removing leakage errors. The next round of error detection proceeds with data qubits $q_2q_4q_6$ and measurement qubits $q_3q_5$ followed by data qubit teleportation to the qubit above, after which the entire cycle repeats.

\begin{figure}
\begin{center}
\resizebox{85mm}{!}{\includegraphics[viewport=60 60 545 430, clip=true]{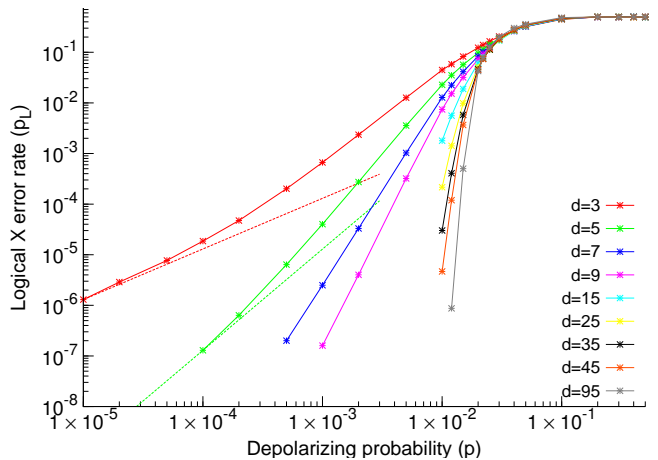}}
\end{center}
\caption{(Color online) probability of logical $X$ error $p_L$ as a function of the depolarizing error probability $p$ and leakage error probability $0.1p$ for a range of distances $d=3\ldots 95$. Referring to the left of the figure, the distance increases top to bottom. The circuit of Fig.~\ref{leak_recover} has been used on each data qubit after each round of error detection to remove leakage errors.}\label{logx_tel}
\end{figure}

The performance of this alternating error detection and teleportation cycle is shown in Fig.~\ref{logx_tel}. It can be seen that while a high power of $p$ suppression of logical error is observed for high distances, a significant improvement on Fig.~\ref{logx_loss}, the suppression is only linear at distance 3 and quadratic at distance 5. This is due to leakage errors on measurement qubits, which can corrupt both neighboring data qubits, creating a two-qubit error chain out of a single leakage error. This means that the logical error rate will be proportional to $\lceil d/4 \rceil$ at low $p$, not proportional to $\lceil d/2 \rceil$ as was observed in Fig.~\ref{logx} without loss.

\section{Conclusion}
\label{conc}

Long-lived leakage is highly detrimental to the performance of topological codes, making logical error rates linear in $p$ at arbitrary code distances, however this does not prevent the arbitrary suppression of logical error as even a very long-lived leakage event is still a local error in time and hence correctable using topological techniques. It is highly advisable to use hardware techniques to periodically remove leakage from the system rather than rely on the weak natural suppression of leakage errors inherent in a topological code as even very low probability leakage, if long-lived, will lead to logical errors with high probability unless an excessive number of qubits are used.

The teleportation circuit of Fig.~\ref{leak_recover}, used once per data qubit per error detection cycle, improves logical error suppression to $O(p^{\lceil d/4 \rceil})$. One could in principle insert a teleportation step in the middle of the stabilizer measurement procedure to remove spatially correlated errors arising from leakage, however it is possible, and desirable, that a deeper understanding of the physical leakage process in specific systems could lead to better and simpler techniques restoring the theoretical maximum $O(p^{\lceil d/2 \rceil})$ suppression. We shall search for such techniques for superconducting qubits in future work.

\section{Acknowledgements}
\label{ack}

This research was funded by the US Office of the Director of National Intelligence (ODNI), Intelligence Advanced Research Projects Activity (IARPA), through the US Army Research Office grant No. W911NF-10-1-0334. Supported in part by the Australian Research Council Centre of Excellence for Quantum Computation and Communication Technology (project number CE110001027) and the US National Security Agency and the US Army Research Office under contract number W911NF-13-1-0024. All statements of fact, opinion or conclusions contained herein are those of the authors and should not be construed as representing the official views or policies of IARPA, the ODNI, or the US Government.

\bibliography{../References}

\begin{thebibliography}{39}%
\makeatletter
\providecommand \@ifxundefined [1]{%
 \@ifx{#1\undefined}
}%
\providecommand \@ifnum [1]{%
 \ifnum #1\expandafter \@firstoftwo
 \else \expandafter \@secondoftwo
 \fi
}%
\providecommand \@ifx [1]{%
 \ifx #1\expandafter \@firstoftwo
 \else \expandafter \@secondoftwo
 \fi
}%
\providecommand \natexlab [1]{#1}%
\providecommand \enquote  [1]{``#1''}%
\providecommand \bibnamefont  [1]{#1}%
\providecommand \bibfnamefont [1]{#1}%
\providecommand \citenamefont [1]{#1}%
\providecommand \href@noop [0]{\@secondoftwo}%
\providecommand \href [0]{\begingroup \@sanitize@url \@href}%
\providecommand \@href[1]{\@@startlink{#1}\@@href}%
\providecommand \@@href[1]{\endgroup#1\@@endlink}%
\providecommand \@sanitize@url [0]{\catcode `\\12\catcode `\$12\catcode
  `\&12\catcode `\#12\catcode `\^12\catcode `\_12\catcode `\%12\relax}%
\providecommand \@@startlink[1]{}%
\providecommand \@@endlink[0]{}%
\providecommand \url  [0]{\begingroup\@sanitize@url \@url }%
\providecommand \@url [1]{\endgroup\@href {#1}{\urlprefix }}%
\providecommand \urlprefix  [0]{URL }%
\providecommand \Eprint [0]{\href }%
\providecommand \doibase [0]{http://dx.doi.org/}%
\providecommand \selectlanguage [0]{\@gobble}%
\providecommand \bibinfo  [0]{\@secondoftwo}%
\providecommand \bibfield  [0]{\@secondoftwo}%
\providecommand \translation [1]{[#1]}%
\providecommand \BibitemOpen [0]{}%
\providecommand \bibitemStop [0]{}%
\providecommand \bibitemNoStop [0]{.\EOS\space}%
\providecommand \EOS [0]{\spacefactor3000\relax}%
\providecommand \BibitemShut  [1]{\csname bibitem#1\endcsname}%
\let\auto@bib@innerbib\@empty
\bibitem [{\citenamefont {Knill}\ \emph {et~al.}(1996)\citenamefont {Knill},
  \citenamefont {Laflamme},\ and\ \citenamefont {Zurek}}]{Knil96b}%
  \BibitemOpen
  \bibfield  {author} {\bibinfo {author} {\bibfnamefont {E.}~\bibnamefont
  {Knill}}, \bibinfo {author} {\bibfnamefont {R.}~\bibnamefont {Laflamme}}, \
  and\ \bibinfo {author} {\bibfnamefont {W.~H.}\ \bibnamefont {Zurek}},\
  }\href@noop {} {\emph {\bibinfo {title} {Accuracy Threshold for Quantum
  Computation}}},\ \bibinfo {type} {Tech. Rep.}\ \bibinfo {number}
  {LAUR-96-2199}\ (\bibinfo  {institution} {Los Alamos National Laboratory},\
  \bibinfo {year} {1996})\ \bibinfo {note} {quant-ph/9610011}\BibitemShut
  {NoStop}%
\bibitem [{\citenamefont {Aharonov}\ and\ \citenamefont
  {Ben-Or}(1997)}]{Ahar97}%
  \BibitemOpen
  \bibfield  {author} {\bibinfo {author} {\bibfnamefont {D.}~\bibnamefont
  {Aharonov}}\ and\ \bibinfo {author} {\bibfnamefont {M.}~\bibnamefont
  {Ben-Or}},\ }\href@noop {} {\bibfield  {journal} {\bibinfo  {journal} {Proc.
  ACM STOC}\ }\textbf {\bibinfo {volume} {29}},\ \bibinfo {pages} {176}
  (\bibinfo {year} {1997})},\ \bibinfo {note} {quant-ph/9611025}\BibitemShut
  {NoStop}%
\bibitem [{\citenamefont {Kitaev}(1997)}]{Kita97}%
  \BibitemOpen
  \bibfield  {author} {\bibinfo {author} {\bibfnamefont {A.~Y.}\ \bibnamefont
  {Kitaev}},\ }\href@noop {} {\bibfield  {journal} {\bibinfo  {journal} {Russ.
  Math. Surv.}\ }\textbf {\bibinfo {volume} {52}},\ \bibinfo {pages} {1191}
  (\bibinfo {year} {1997})}\BibitemShut {NoStop}%
\bibitem [{\citenamefont {Aliferis}\ and\ \citenamefont
  {Terhal}(2007)}]{Alif07d}%
  \BibitemOpen
  \bibfield  {author} {\bibinfo {author} {\bibfnamefont {P.}~\bibnamefont
  {Aliferis}}\ and\ \bibinfo {author} {\bibfnamefont {B.~M.}\ \bibnamefont
  {Terhal}},\ }\href@noop {} {\bibfield  {journal} {\bibinfo  {journal} {Quant.
  Info. Comp.}\ }\textbf {\bibinfo {volume} {7}},\ \bibinfo {pages} {139}
  (\bibinfo {year} {2007})},\ \bibinfo {note} {quant-ph/0511065}\BibitemShut
  {NoStop}%
\bibitem [{\citenamefont {Shor}(1995)}]{Shor95}%
  \BibitemOpen
  \bibfield  {author} {\bibinfo {author} {\bibfnamefont {P.~W.}\ \bibnamefont
  {Shor}},\ }\href@noop {} {\bibfield  {journal} {\bibinfo  {journal} {Phys.
  Rev. A}\ }\textbf {\bibinfo {volume} {52}},\ \bibinfo {pages} {R2493}
  (\bibinfo {year} {1995})}\BibitemShut {NoStop}%
\bibitem [{\citenamefont {Calderbank}\ and\ \citenamefont
  {Shor}(1996)}]{Cald95}%
  \BibitemOpen
  \bibfield  {author} {\bibinfo {author} {\bibfnamefont {A.~R.}\ \bibnamefont
  {Calderbank}}\ and\ \bibinfo {author} {\bibfnamefont {P.~W.}\ \bibnamefont
  {Shor}},\ }\href@noop {} {\bibfield  {journal} {\bibinfo  {journal} {Phys.
  Rev. A}\ }\textbf {\bibinfo {volume} {54}},\ \bibinfo {pages} {1098}
  (\bibinfo {year} {1996})},\ \bibinfo {note} {quant-ph/9512032}\BibitemShut
  {NoStop}%
\bibitem [{\citenamefont {Steane}(1996)}]{Stea96}%
  \BibitemOpen
  \bibfield  {author} {\bibinfo {author} {\bibfnamefont {A.~M.}\ \bibnamefont
  {Steane}},\ }\href@noop {} {\bibfield  {journal} {\bibinfo  {journal} {Proc.
  R. Soc. Lond. A}\ }\textbf {\bibinfo {volume} {452}},\ \bibinfo {pages}
  {2551} (\bibinfo {year} {1996})},\ \bibinfo {note}
  {quant-ph/9601029}\BibitemShut {NoStop}%
\bibitem [{\citenamefont {Knill}(2005)}]{Knil04c}%
  \BibitemOpen
  \bibfield  {author} {\bibinfo {author} {\bibfnamefont {E.}~\bibnamefont
  {Knill}},\ }\href@noop {} {\bibfield  {journal} {\bibinfo  {journal}
  {Nature}\ }\textbf {\bibinfo {volume} {434}},\ \bibinfo {pages} {39}
  (\bibinfo {year} {2005})},\ \bibinfo {note} {quant-ph/0410199}\BibitemShut
  {NoStop}%
\bibitem [{\citenamefont {Bacon}(2006)}]{Baco06}%
  \BibitemOpen
  \bibfield  {author} {\bibinfo {author} {\bibfnamefont {D.}~\bibnamefont
  {Bacon}},\ }\href@noop {} {\bibfield  {journal} {\bibinfo  {journal} {Phys.
  Rev. A}\ }\textbf {\bibinfo {volume} {73}},\ \bibinfo {pages} {012340}
  (\bibinfo {year} {2006})},\ \bibinfo {note} {quant-ph/0506023}\BibitemShut
  {NoStop}%
\bibitem [{\citenamefont {Bravyi}\ and\ \citenamefont {Kitaev}(1998)}]{Brav98}%
  \BibitemOpen
  \bibfield  {author} {\bibinfo {author} {\bibfnamefont {S.~B.}\ \bibnamefont
  {Bravyi}}\ and\ \bibinfo {author} {\bibfnamefont {A.~Y.}\ \bibnamefont
  {Kitaev}},\ }\href@noop {} {\bibfield  {journal} {\bibinfo  {journal}
  {quant-ph/9811052}\ } (\bibinfo {year} {1998})}\BibitemShut {NoStop}%
\bibitem [{\citenamefont {Dennis}\ \emph {et~al.}(2002)\citenamefont {Dennis},
  \citenamefont {Kitaev}, \citenamefont {Landahl},\ and\ \citenamefont
  {Preskill}}]{Denn02}%
  \BibitemOpen
  \bibfield  {author} {\bibinfo {author} {\bibfnamefont {E.}~\bibnamefont
  {Dennis}}, \bibinfo {author} {\bibfnamefont {A.}~\bibnamefont {Kitaev}},
  \bibinfo {author} {\bibfnamefont {A.}~\bibnamefont {Landahl}}, \ and\
  \bibinfo {author} {\bibfnamefont {J.}~\bibnamefont {Preskill}},\ }\href@noop
  {} {\bibfield  {journal} {\bibinfo  {journal} {J. Math. Phys.}\ }\textbf
  {\bibinfo {volume} {43}},\ \bibinfo {pages} {4452} (\bibinfo {year}
  {2002})},\ \bibinfo {note} {quant-ph/0110143}\BibitemShut {NoStop}%
\bibitem [{\citenamefont {Bombin}\ and\ \citenamefont
  {Martin-Delgado}(2006)}]{Bomb06}%
  \BibitemOpen
  \bibfield  {author} {\bibinfo {author} {\bibfnamefont {H.}~\bibnamefont
  {Bombin}}\ and\ \bibinfo {author} {\bibfnamefont {M.~A.}\ \bibnamefont
  {Martin-Delgado}},\ }\href@noop {} {\bibfield  {journal} {\bibinfo  {journal}
  {Phys. Rev. Lett.}\ }\textbf {\bibinfo {volume} {97}},\ \bibinfo {pages}
  {180501} (\bibinfo {year} {2006})},\ \bibinfo {note}
  {quant-ph/0605138}\BibitemShut {NoStop}%
\bibitem [{\citenamefont {Raussendorf}\ and\ \citenamefont
  {Harrington}(2007)}]{Raus07}%
  \BibitemOpen
  \bibfield  {author} {\bibinfo {author} {\bibfnamefont {R.}~\bibnamefont
  {Raussendorf}}\ and\ \bibinfo {author} {\bibfnamefont {J.}~\bibnamefont
  {Harrington}},\ }\href@noop {} {\bibfield  {journal} {\bibinfo  {journal}
  {Phys. Rev. Lett.}\ }\textbf {\bibinfo {volume} {98}},\ \bibinfo {pages}
  {190504} (\bibinfo {year} {2007})},\ \bibinfo {note}
  {quant-ph/0610082}\BibitemShut {NoStop}%
\bibitem [{\citenamefont {Raussendorf}\ \emph {et~al.}(2007)\citenamefont
  {Raussendorf}, \citenamefont {Harrington},\ and\ \citenamefont
  {Goyal}}]{Raus07d}%
  \BibitemOpen
  \bibfield  {author} {\bibinfo {author} {\bibfnamefont {R.}~\bibnamefont
  {Raussendorf}}, \bibinfo {author} {\bibfnamefont {J.}~\bibnamefont
  {Harrington}}, \ and\ \bibinfo {author} {\bibfnamefont {K.}~\bibnamefont
  {Goyal}},\ }\href@noop {} {\bibfield  {journal} {\bibinfo  {journal} {New J.
  Phys.}\ }\textbf {\bibinfo {volume} {9}},\ \bibinfo {pages} {199} (\bibinfo
  {year} {2007})},\ \bibinfo {note} {quant-ph/0703143}\BibitemShut {NoStop}%
\bibitem [{\citenamefont {Fowler}\ and\ \citenamefont {Goyal}(2009)}]{Fowl09}%
  \BibitemOpen
  \bibfield  {author} {\bibinfo {author} {\bibfnamefont {A.~G.}\ \bibnamefont
  {Fowler}}\ and\ \bibinfo {author} {\bibfnamefont {K.}~\bibnamefont {Goyal}},\
  }\href@noop {} {\bibfield  {journal} {\bibinfo  {journal} {Quant. Info.
  Comput.}\ }\textbf {\bibinfo {volume} {9}},\ \bibinfo {pages} {721} (\bibinfo
  {year} {2009})},\ \bibinfo {note} {arXiv:0805.3202}\BibitemShut {NoStop}%
\bibitem [{\citenamefont {Ohzeki}(2009)}]{Ohze09b}%
  \BibitemOpen
  \bibfield  {author} {\bibinfo {author} {\bibfnamefont {M.}~\bibnamefont
  {Ohzeki}},\ }\href@noop {} {\bibfield  {journal} {\bibinfo  {journal} {Phys.
  Rev. E}\ }\textbf {\bibinfo {volume} {80}},\ \bibinfo {pages} {011141}
  (\bibinfo {year} {2009})},\ \bibinfo {note} {arXiv:0903.2102}\BibitemShut
  {NoStop}%
\bibitem [{\citenamefont {Katzgraber}\ \emph {et~al.}(2010)\citenamefont
  {Katzgraber}, \citenamefont {Bombin}, \citenamefont {Andrist},\ and\
  \citenamefont {Martin-Delgado}}]{Katz10}%
  \BibitemOpen
  \bibfield  {author} {\bibinfo {author} {\bibfnamefont {H.~G.}\ \bibnamefont
  {Katzgraber}}, \bibinfo {author} {\bibfnamefont {H.}~\bibnamefont {Bombin}},
  \bibinfo {author} {\bibfnamefont {R.~S.}\ \bibnamefont {Andrist}}, \ and\
  \bibinfo {author} {\bibfnamefont {M.~A.}\ \bibnamefont {Martin-Delgado}},\
  }\href@noop {} {\bibfield  {journal} {\bibinfo  {journal} {Phys. Rev. A}\
  }\textbf {\bibinfo {volume} {81}},\ \bibinfo {pages} {012319} (\bibinfo
  {year} {2010})},\ \bibinfo {note} {arXiv:0910.0573}\BibitemShut {NoStop}%
\bibitem [{\citenamefont {Bombin}(2011)}]{Bomb10}%
  \BibitemOpen
  \bibfield  {author} {\bibinfo {author} {\bibfnamefont {H.}~\bibnamefont
  {Bombin}},\ }\href@noop {} {\bibfield  {journal} {\bibinfo  {journal} {New J.
  Phys.}\ }\textbf {\bibinfo {volume} {13}},\ \bibinfo {pages} {043005}
  (\bibinfo {year} {2011})},\ \bibinfo {note} {arXiv:1006.5260}\BibitemShut
  {NoStop}%
\bibitem [{\citenamefont {Fowler}(2011)}]{Fowl11}%
  \BibitemOpen
  \bibfield  {author} {\bibinfo {author} {\bibfnamefont {A.~G.}\ \bibnamefont
  {Fowler}},\ }\href@noop {} {\bibfield  {journal} {\bibinfo  {journal} {Phys.
  Rev. A}\ }\textbf {\bibinfo {volume} {83}},\ \bibinfo {pages} {042310}
  (\bibinfo {year} {2011})},\ \bibinfo {note} {arXiv:0806.4827}\BibitemShut
  {NoStop}%
\bibitem [{\citenamefont {Fowler}\ \emph
  {et~al.}(2012{\natexlab{a}})\citenamefont {Fowler}, \citenamefont
  {Mariantoni}, \citenamefont {Martinis},\ and\ \citenamefont
  {Cleland}}]{Fowl12f}%
  \BibitemOpen
  \bibfield  {author} {\bibinfo {author} {\bibfnamefont {A.~G.}\ \bibnamefont
  {Fowler}}, \bibinfo {author} {\bibfnamefont {M.}~\bibnamefont {Mariantoni}},
  \bibinfo {author} {\bibfnamefont {J.~M.}\ \bibnamefont {Martinis}}, \ and\
  \bibinfo {author} {\bibfnamefont {A.~N.}\ \bibnamefont {Cleland}},\
  }\href@noop {} {\bibfield  {journal} {\bibinfo  {journal} {Phys. Rev. A}\
  }\textbf {\bibinfo {volume} {86}},\ \bibinfo {pages} {032324} (\bibinfo
  {year} {2012}{\natexlab{a}})},\ \bibinfo {note} {arXiv:1208.0928}\BibitemShut
  {NoStop}%
\bibitem [{\citenamefont {Harrington}(2004)}]{Harr04}%
  \BibitemOpen
  \bibfield  {author} {\bibinfo {author} {\bibfnamefont {J.~W.}\ \bibnamefont
  {Harrington}},\ }\emph {\bibinfo {title} {Analysis of quantum
  error-correcting codes: symplectic lattice codes and toric codes}},\
  \href@noop {} {Ph.D. thesis},\ \bibinfo  {school} {California Institute of
  Technology, Pasadena, California} (\bibinfo {year} {2004})\BibitemShut
  {NoStop}%
\bibitem [{\citenamefont {Bravyi}\ and\ \citenamefont {Haah}(2011)}]{Brav11}%
  \BibitemOpen
  \bibfield  {author} {\bibinfo {author} {\bibfnamefont {S.}~\bibnamefont
  {Bravyi}}\ and\ \bibinfo {author} {\bibfnamefont {J.}~\bibnamefont {Haah}},\
  }\href@noop {} {\bibfield  {journal} {\bibinfo  {journal} {arXiv:1112.3252}\
  } (\bibinfo {year} {2011})}\BibitemShut {NoStop}%
\bibitem [{\citenamefont {Fowler}(2012)}]{Fowl12e}%
  \BibitemOpen
  \bibfield  {author} {\bibinfo {author} {\bibfnamefont {A.~G.}\ \bibnamefont
  {Fowler}},\ }\href@noop {} {\bibfield  {journal} {\bibinfo  {journal} {Phys.
  Rev. Lett.}\ }\textbf {\bibinfo {volume} {109}},\ \bibinfo {pages} {180502}
  (\bibinfo {year} {2012})},\ \bibinfo {note} {arXiv:1206.0800}\BibitemShut
  {NoStop}%
\bibitem [{\citenamefont {Cirac}\ and\ \citenamefont {Zoller}(1995)}]{Cira95}%
  \BibitemOpen
  \bibfield  {author} {\bibinfo {author} {\bibfnamefont {J.~I.}\ \bibnamefont
  {Cirac}}\ and\ \bibinfo {author} {\bibfnamefont {P.}~\bibnamefont {Zoller}},\
  }\href@noop {} {\bibfield  {journal} {\bibinfo  {journal} {Phys. Rev. Lett.}\
  }\textbf {\bibinfo {volume} {74}},\ \bibinfo {pages} {4091} (\bibinfo {year}
  {1995})}\BibitemShut {NoStop}%
\bibitem [{\citenamefont {Byrd}\ \emph {et~al.}(2005)\citenamefont {Byrd},
  \citenamefont {Lidar}, \citenamefont {Wu},\ and\ \citenamefont
  {Zanardi}}]{Byrd05}%
  \BibitemOpen
  \bibfield  {author} {\bibinfo {author} {\bibfnamefont {M.~S.}\ \bibnamefont
  {Byrd}}, \bibinfo {author} {\bibfnamefont {D.~A.}\ \bibnamefont {Lidar}},
  \bibinfo {author} {\bibfnamefont {L.-A.}\ \bibnamefont {Wu}}, \ and\ \bibinfo
  {author} {\bibfnamefont {P.}~\bibnamefont {Zanardi}},\ }\href@noop {}
  {\bibfield  {journal} {\bibinfo  {journal} {Phys. Rev. A}\ }\textbf {\bibinfo
  {volume} {71}},\ \bibinfo {pages} {052301} (\bibinfo {year} {2005})},\
  \bibinfo {note} {quant-ph/0409049}\BibitemShut {NoStop}%
\bibitem [{\citenamefont {Fong}\ and\ \citenamefont {Wandzura}(2011)}]{Fong11}%
  \BibitemOpen
  \bibfield  {author} {\bibinfo {author} {\bibfnamefont {B.~H.}\ \bibnamefont
  {Fong}}\ and\ \bibinfo {author} {\bibfnamefont {S.~M.}\ \bibnamefont
  {Wandzura}},\ }\href@noop {} {\bibfield  {journal} {\bibinfo  {journal}
  {Quant. Info. Comp.}\ }\textbf {\bibinfo {volume} {11}},\ \bibinfo {pages}
  {1003} (\bibinfo {year} {2011})},\ \bibinfo {note}
  {arXiv:1102.2909}\BibitemShut {NoStop}%
\bibitem [{\citenamefont {Zhou}\ \emph {et~al.}(2005)\citenamefont {Zhou},
  \citenamefont {Chu},\ and\ \citenamefont {Han}}]{Zhou05}%
  \BibitemOpen
  \bibfield  {author} {\bibinfo {author} {\bibfnamefont {Z.}~\bibnamefont
  {Zhou}}, \bibinfo {author} {\bibfnamefont {S.-I.}\ \bibnamefont {Chu}}, \
  and\ \bibinfo {author} {\bibfnamefont {S.}~\bibnamefont {Han}},\ }\href@noop
  {} {\bibfield  {journal} {\bibinfo  {journal} {Phys. Rev. Lett.}\ }\textbf
  {\bibinfo {volume} {95}},\ \bibinfo {pages} {120501} (\bibinfo {year}
  {2005})},\ \bibinfo {note} {quant-ph/0506257}\BibitemShut {NoStop}%
\bibitem [{\citenamefont {Wu}\ \emph {et~al.}(2006)\citenamefont {Wu},
  \citenamefont {Liu},\ and\ \citenamefont {Li}}]{Tao06}%
  \BibitemOpen
  \bibfield  {author} {\bibinfo {author} {\bibfnamefont {T.}~\bibnamefont
  {Wu}}, \bibinfo {author} {\bibfnamefont {J.}~\bibnamefont {Liu}}, \ and\
  \bibinfo {author} {\bibfnamefont {Z.}~\bibnamefont {Li}},\ }\href@noop {}
  {\bibfield  {journal} {\bibinfo  {journal} {Chin. Phys. Lett.}\ }\textbf
  {\bibinfo {volume} {23}},\ \bibinfo {pages} {971} (\bibinfo {year} {2006})},\
  \bibinfo {note} {cond-mat/0511012}\BibitemShut {NoStop}%
\bibitem [{\citenamefont {Motzoi}\ \emph {et~al.}(2009)\citenamefont {Motzoi},
  \citenamefont {Gambetta}, \citenamefont {Rebentrost},\ and\ \citenamefont
  {Wilhelm}}]{Motz09}%
  \BibitemOpen
  \bibfield  {author} {\bibinfo {author} {\bibfnamefont {F.}~\bibnamefont
  {Motzoi}}, \bibinfo {author} {\bibfnamefont {J.~M.}\ \bibnamefont
  {Gambetta}}, \bibinfo {author} {\bibfnamefont {P.}~\bibnamefont
  {Rebentrost}}, \ and\ \bibinfo {author} {\bibfnamefont {F.~K.}\ \bibnamefont
  {Wilhelm}},\ }\href@noop {} {\bibfield  {journal} {\bibinfo  {journal} {Phys.
  Rev. Lett.}\ }\textbf {\bibinfo {volume} {103}},\ \bibinfo {pages} {110501}
  (\bibinfo {year} {2009})},\ \bibinfo {note} {arXiv:0901.0534}\BibitemShut
  {NoStop}%
\bibitem [{\citenamefont {Ferron}\ and\ \citenamefont
  {Dominguez}(2010)}]{Ferr10}%
  \BibitemOpen
  \bibfield  {author} {\bibinfo {author} {\bibfnamefont {A.}~\bibnamefont
  {Ferron}}\ and\ \bibinfo {author} {\bibfnamefont {D.}~\bibnamefont
  {Dominguez}},\ }\href@noop {} {\bibfield  {journal} {\bibinfo  {journal}
  {Phys. Rev. B}\ }\textbf {\bibinfo {volume} {81}},\ \bibinfo {pages} {104505}
  (\bibinfo {year} {2010})},\ \bibinfo {note} {arXiv:0910.5640}\BibitemShut
  {NoStop}%
\bibitem [{\citenamefont {Herrera-Martí}\ \emph {et~al.}(2012)\citenamefont
  {Herrera-Martí}, \citenamefont {Nazir},\ and\ \citenamefont
  {Barrett}}]{Herr12b}%
  \BibitemOpen
  \bibfield  {author} {\bibinfo {author} {\bibfnamefont {D.~A.}\ \bibnamefont
  {Herrera-Martí}}, \bibinfo {author} {\bibfnamefont {A.}~\bibnamefont
  {Nazir}}, \ and\ \bibinfo {author} {\bibfnamefont {S.~D.}\ \bibnamefont
  {Barrett}},\ }\href@noop {} {\bibfield  {journal} {\bibinfo  {journal}
  {arXiv:1212.4557}\ } (\bibinfo {year} {2012})}\BibitemShut {NoStop}%
\bibitem [{\citenamefont {Ghosh}\ \emph {et~al.}(2013)\citenamefont {Ghosh},
  \citenamefont {Fowler}, \citenamefont {Martinis},\ and\ \citenamefont
  {Geller}}]{Ghos13}%
  \BibitemOpen
  \bibfield  {author} {\bibinfo {author} {\bibfnamefont {J.}~\bibnamefont
  {Ghosh}}, \bibinfo {author} {\bibfnamefont {A.~G.}\ \bibnamefont {Fowler}},
  \bibinfo {author} {\bibfnamefont {J.~M.}\ \bibnamefont {Martinis}}, \ and\
  \bibinfo {author} {\bibfnamefont {M.~R.}\ \bibnamefont {Geller}},\
  }\href@noop {} {\bibfield  {journal} {\bibinfo  {journal} {arXiv:1306.0925}\
  } (\bibinfo {year} {2013})}\BibitemShut {NoStop}%
\bibitem [{\citenamefont {Xu}\ and\ \citenamefont {Wan}(2008)}]{Xu02}%
  \BibitemOpen
  \bibfield  {author} {\bibinfo {author} {\bibfnamefont {H.}~\bibnamefont
  {Xu}}\ and\ \bibinfo {author} {\bibfnamefont {X.}~\bibnamefont {Wan}},\
  }\href@noop {} {\bibfield  {journal} {\bibinfo  {journal} {Phys. Rev. A}\
  }\textbf {\bibinfo {volume} {78}},\ \bibinfo {pages} {042325} (\bibinfo
  {year} {2008})},\ \bibinfo {note} {arXiv:0802.4213}\BibitemShut {NoStop}%
\bibitem [{\citenamefont {Ainsworth}\ and\ \citenamefont
  {Slingerland}(2011)}]{Ains11}%
  \BibitemOpen
  \bibfield  {author} {\bibinfo {author} {\bibfnamefont {R.}~\bibnamefont
  {Ainsworth}}\ and\ \bibinfo {author} {\bibfnamefont {J.~K.}\ \bibnamefont
  {Slingerland}},\ }\href@noop {} {\bibfield  {journal} {\bibinfo  {journal}
  {New J. Phys.}\ }\textbf {\bibinfo {volume} {13}},\ \bibinfo {pages} {065030}
  (\bibinfo {year} {2011})},\ \bibinfo {note} {arXiv:1102.5029}\BibitemShut
  {NoStop}%
\bibitem [{\citenamefont {Fowler}\ \emph
  {et~al.}(2012{\natexlab{b}})\citenamefont {Fowler}, \citenamefont
  {Whiteside}, \citenamefont {McInnes},\ and\ \citenamefont
  {Rabbani}}]{Fowl12d}%
  \BibitemOpen
  \bibfield  {author} {\bibinfo {author} {\bibfnamefont {A.~G.}\ \bibnamefont
  {Fowler}}, \bibinfo {author} {\bibfnamefont {A.~C.}\ \bibnamefont
  {Whiteside}}, \bibinfo {author} {\bibfnamefont {A.~L.}\ \bibnamefont
  {McInnes}}, \ and\ \bibinfo {author} {\bibfnamefont {A.}~\bibnamefont
  {Rabbani}},\ }\href@noop {} {\bibfield  {journal} {\bibinfo  {journal} {Phys.
  Rev. X}\ }\textbf {\bibinfo {volume} {2}},\ \bibinfo {pages} {041003}
  (\bibinfo {year} {2012}{\natexlab{b}})},\ \bibinfo {note} {arXiv:1202.6111,
  http://topqec.com.au/autotune.html}\BibitemShut {NoStop}%
\bibitem [{\citenamefont {Edmonds}(1965{\natexlab{a}})}]{Edmo65a}%
  \BibitemOpen
  \bibfield  {author} {\bibinfo {author} {\bibfnamefont {J.}~\bibnamefont
  {Edmonds}},\ }\href@noop {} {\bibfield  {journal} {\bibinfo  {journal}
  {Canad. J. Math.}\ }\textbf {\bibinfo {volume} {17}},\ \bibinfo {pages} {449}
  (\bibinfo {year} {1965}{\natexlab{a}})}\BibitemShut {NoStop}%
\bibitem [{\citenamefont {Edmonds}(1965{\natexlab{b}})}]{Edmo65b}%
  \BibitemOpen
  \bibfield  {author} {\bibinfo {author} {\bibfnamefont {J.}~\bibnamefont
  {Edmonds}},\ }\href@noop {} {\bibfield  {journal} {\bibinfo  {journal} {J.
  Res. Nat. Bur. Standards}\ }\textbf {\bibinfo {volume} {69B}},\ \bibinfo
  {pages} {125} (\bibinfo {year} {1965}{\natexlab{b}})}\BibitemShut {NoStop}%
\bibitem [{\citenamefont {Fowler}\ \emph
  {et~al.}(2012{\natexlab{c}})\citenamefont {Fowler}, \citenamefont
  {Whiteside},\ and\ \citenamefont {Hollenberg}}]{Fowl11b}%
  \BibitemOpen
  \bibfield  {author} {\bibinfo {author} {\bibfnamefont {A.~G.}\ \bibnamefont
  {Fowler}}, \bibinfo {author} {\bibfnamefont {A.~C.}\ \bibnamefont
  {Whiteside}}, \ and\ \bibinfo {author} {\bibfnamefont {L.~C.~L.}\
  \bibnamefont {Hollenberg}},\ }\href@noop {} {\bibfield  {journal} {\bibinfo
  {journal} {Phys. Rev. Lett.}\ }\textbf {\bibinfo {volume} {108}},\ \bibinfo
  {pages} {180501} (\bibinfo {year} {2012}{\natexlab{c}})},\ \bibinfo {note}
  {arXiv:1110.5133}\BibitemShut {NoStop}%
\bibitem [{\citenamefont {Fowler}\ \emph
  {et~al.}(2012{\natexlab{d}})\citenamefont {Fowler}, \citenamefont
  {Whiteside},\ and\ \citenamefont {Hollenberg}}]{Fowl12c}%
  \BibitemOpen
  \bibfield  {author} {\bibinfo {author} {\bibfnamefont {A.~G.}\ \bibnamefont
  {Fowler}}, \bibinfo {author} {\bibfnamefont {A.~C.}\ \bibnamefont
  {Whiteside}}, \ and\ \bibinfo {author} {\bibfnamefont {L.~C.~L.}\
  \bibnamefont {Hollenberg}},\ }\href@noop {} {\bibfield  {journal} {\bibinfo
  {journal} {Phys. Rev. A}\ }\textbf {\bibinfo {volume} {86}},\ \bibinfo
  {pages} {042313} (\bibinfo {year} {2012}{\natexlab{d}})},\ \bibinfo {note}
  {arXiv:1202.5602}\BibitemShut {NoStop}%
\end{thebibliography}%

\end{document}